\documentclass[11pt,a4paper]{article}
\pdfoutput=1
\usepackage{jheppub}
\usepackage[utf8]{inputenc}

\usepackage{color}
\usepackage{amsmath}
\usepackage{mathtools}
\usepackage{esint}
\usepackage{pifont}
\usepackage{bbold}

\usepackage{bbm}
\usepackage{verbatim}   
\usepackage{subfigure}
\usepackage{acronym}

\usepackage{amsfonts}
\usepackage{amssymb}
\usepackage{mathrsfs}
\usepackage{graphicx}
\usepackage{multirow}
 \usepackage{slashed}

\usepackage{caption}

\usepackage{soul}

\makeatletter
\def\@fpheader{\relax}
\makeatother

\definecolor{mypurple}{RGB}{164,64,214}

\title{Discrete Gauge Symmetries and the Weak Gravity Conjecture}

\author[a]{Nathaniel Craig,}
\emailAdd{ncraig@physics.ucsb.edu}
\author[b]{Isabel Garcia Garcia}
\emailAdd{isabel@kitp.ucsb.edu}
\author[a]{and Seth Koren}
\emailAdd{koren@physics.ucsb.edu}

\affiliation[a]{Department of Physics, University of California, Santa Barbara, CA 93106, USA}
\affiliation[b]{Kavli Institute for Theoretical Physics, University of California, Santa Barbara, CA 93106, USA}

\abstract{
In theories with discrete Abelian gauge groups, requiring that black holes be able to lose their charge as they evaporate leads to an upper bound on the product of a charged particle's mass and the cutoff scale above which the effective description of the theory breaks down. This suggests that a non-trivial version of the Weak Gravity Conjecture (WGC) may also apply to gauge symmetries that are discrete, despite there being no associated massless field, therefore pushing the conjecture beyond the slogan that `gravity is the weakest force'. Here, we take a step towards making this expectation more precise by studying $\mathbb{Z}_N$ and $\mathbb{Z}_2^N$ gauge symmetries realised via theories of spontaneous symmetry breaking. We show that applying the WGC to a dual description of an Abelian Higgs model leads to constraints that allow us to saturate but not violate existing bounds on discrete symmetries based on black hole arguments. In this setting, considering the effect of discrete hair on black holes naturally identifies the cutoff of the effective theory with the scale of spontaneous symmetry breaking, and provides a mechanism through which discrete hair can be lost without modifying the gravitational sector. We explore the possible implications of these arguments for understanding the smallness of the weak scale compared to $M_{Pl}$.
}

\begin{document}

\maketitle

\section{Introduction}

The Weak Gravity Conjecture (WGC) \cite{ArkaniHamed:2006dz} states that a $U(1)$ gauge theory consistently coupled to gravity must contain a particle with mass $m$ and charge $g$ satisfying
\begin{equation}
	m \leq \frac{g M_{Pl}}{\sqrt{4 \pi}} \ .
\label{eq:WGC}
\end{equation}
In what follows we ignore the factor of $\sqrt{4 \pi}$ and use $m \lesssim g M_{Pl}$, as $\mathcal{O}(1)$ refinements have no bearing on our discussion.
Because this charge-to-mass-ratio is larger than that of an extremal black hole (BH), the corresponding state is referred to as being `super-extremal'.
Since the electrostatic force between two such particles overcomes their gravitational attraction, the WGC can be recast as the statement that `gravity is the weakest force'.

Motivation for the conjecture stems from the same type of considerations that question the viability of global symmetries in a theory of quantum gravity, namely the potential presence of a large number of stable remnants at the end of the evaporation process if BHs cannot lose their charge \cite{Abbott:1989jw,Coleman:1989zu,Holman:1992us,Kallosh:1995hi,Banks:2010zn,Susskind:1995da}.
Unlike for global symmetries, for which such number could be infinite, for a gauge theory this number is finite once charge quantization is taken into account, therefore making the problem less severe.
For a $U(1)$ gauge theory with electric charge quantized in units of $g$, the number of different BHs potentially unable to shed their charge is of order $1/g$.
In the regime of weak coupling, $g \ll 1$, this number could be very large, leading to a situation similar in spirit to that for global symmetries.
The presence of an elementary particle satisfying Eq.(\ref{eq:WGC}) allows BHs to lose charge as they evaporate, alleviating the problem.
In particular, it implies that extremal BHs will decay. 
\footnote{It is unclear which version of the WGC should be applied to effective field theories \cite{Saraswat:2016eaz}: In principle, the super-extremal state may not carry the unit of charge. Such considerations, however, will not be relevant for this work, and in the following we stick to the `unit-charge' version of the conjecture for simplicity.}

The same arguments can be applied to BHs carrying magnetic, rather than electric, charge, with the caveat that in a theory containing weakly coupled electrically charged particles we expect magnetic monopoles to be extended, solitonic objects. If the typical size of the monopole solution carrying the unit of magnetic charge is $L \equiv \Lambda^{-1}$, demanding that it satisfies the super-extremality condition leads to an upper bound \cite{ArkaniHamed:2006dz} \footnote{The same result follows by demanding that the unit-charge magnetic monopole is not itself a BH \cite{Saraswat:2016eaz}.}
\begin{equation}
	\Lambda \lesssim g M_{Pl} \ .
\label{eq:WGCmag}
\end{equation}
It was advocated in \cite{ArkaniHamed:2006dz} that $\Lambda$ ought to be interpreted as a cutoff scale beyond which an effective description of the $U(1)$ gauge theory needs to be extended in order to account for the monopoles' structure. For instance, if the $U(1)$ is embedded into a broken non-Abelian gauge group, the scale $\Lambda$ corresponds to the appearance of new degrees of freedom, corresponding to the gauge bosons associated to the broken directions. \footnote{From this point of view, the magnetic scale $\Lambda$ need not have implications stronger than the presence of additional weakly coupled degrees of freedom. This is not in contradiction with stronger variants of the conjecture, such as the `sub-lattice' version of \cite{Heidenreich:2016aqi,Heidenreich:2017sim}, that further indicate a cutoff to four-dimensional QFT at scales of order $g^{1/3} M_{Pl}$ associated to the presence of an infinite tower of states: In the regime of weak coupling $g \ll 1$, the cutoff $g^{1/3} M_{Pl}$ lies parametrically above the scale $g M_{Pl}$.}

The WGC is further supported by the absence of counter-examples in non-trivial string constructions, as well as by additional low energy arguments involving BHs, and an apparent connection to cosmic censorship (see \cite{ArkaniHamed:2006dz,Pal:2010bf,Cheung:2014ega,Harlow:2015lma,Cottrell:2016bty,Hebecker:2017uix,Lee:2018urn,Andriolo:2018lvp,Hamada:2018dde,Bonnefoy:2018mqb,Crisford:2017gsb,Horowitz:2019eum}). There exist many variants of the conjecture \cite{Huang:2006hc,Li:2006jja,Huang:2007st,Cheung:2014vva,Nakayama:2015hga,Heidenreich:2015nta,Heidenreich:2016aqi,Montero:2016tif,Palti:2017elp}, and attempts to find a proof have been the focus of recent literature \cite{Hod:2017uqc,Fisher:2017dbc,Cheung:2018cwt,Montero:2018fns}.
The conjecture further fits in well with two other constraints believed to hold in consistent quantum gravities: namely the absence of exact global symmetries (the WGC precludes a smooth $g \rightarrow 0$ limit), as well as the completeness hypothesis \cite{Polchinski:2003bq}, which requires the presence of dynamical objects carrying all possible charges consistent with the Dirac quantization condition.
Although in general these remain in conjectural form, they have recently been proven in the context of the AdS/CFT correspondence \cite{Harlow:2018jwu}, bringing hope that similar arguments may eventually lead to a proof of some version of the WGC.

The existence of asymptotic observables clearly plays a crucial role in arguments motivating the WGC from infrared considerations involving BH physics as originally discussed in~\cite{ArkaniHamed:2006dz}. For a $U(1)$ gauge theory, the presence of two such observables, electric and magnetic charge, leads to two versions of the conjecture, corresponding to Eq.(\ref{eq:WGC}) and (\ref{eq:WGCmag}) respectively. However, if the $U(1)$ is spontaneously broken, the absence of a long-range classical force a priori precludes similar arguments from surviving in the broken theory. 

An exception to the above statement arises if a discrete subgroup is left unbroken, the simplest example being a $U(1)$ gauge theory broken down to a $\mathbb{Z}_N$ subgroup as a result of the non-zero vacuum expectation value (vev) of a Higgs field carrying charge $g N$, with $N>1$. In the $\mathbb{Z}_N$ phase, electric fields are screened, decaying exponentially away from the source, whereas magnetic flux remains confined into flux tubes, or cosmic strings. In an Abelian Higgs model, the amount of flux confined into strings is quantized in units of $2 \pi / (gN)$, and the tension of the string carrying the unit of flux is $T_s \sim v^2$, with $v$ the Higgs vev. Despite there being no massless fields associated with a discrete gauge symmetry, and therefore no long-range classical force, the $\mathbb{Z}_N$ theory nevertheless features long-range interactions, namely Aharonov-Bohm scattering of cosmic strings with matter \cite{Aharonov:1959fk,Alford:1988sj}. By means of Aharonov-Bohm scattering experiments, discrete electric charge can be measured from far away, and becomes an asymptotic observable.

As a result, BHs may carry discrete charge as hair, a possibility first discussed in \cite{Krauss:1988zc}, with \cite{Alford:1989ch,Alford:1990pt} extending the argument to the non-Abelian case. (Notice this type of hair is purely quantum, crucially depending on both charge quantization and the associated Aharonov-Bohm scattering, and therefore falls outside of the scope of classical no-hair theorems.) BHs carrying discrete electric charge therefore qualify for the same type of thought experiments that led to the WGC for continuous gauge symmetries, as first discussed by Dvali and collaborators~\cite{Dvali:2007hz,Dvali:2007wp,Dvali:2008tq} (see also \cite{Preskill:1990ty}). In a theory with $\mathbb{Z}_N$ gauge group, if the state $\psi$ that carries the unit of discrete charge has mass $m$, then the requirement that there be no remnants left that are stabilized by discrete charge leads to an upper bound relating $m$ and the scale $\Lambda$ beyond which an effective description of the $\mathbb{Z}_N$ gauge theory is expected to no longer be valid, of the form \cite{Dvali:2007hz,Dvali:2008tq}
\begin{equation}
	m \cdot \Lambda \lesssim \frac{M_{Pl}^2}{N} \ .
\label{eq:Dvali}
\end{equation}

Eq.(\ref{eq:Dvali}) also applies if the gauge group is $\mathbb{Z}_2^N$ \cite{Dvali:2007hz,Dvali:2007wp}, under the assumption that the states carrying the unit of electric charge under each of the $\mathbb{Z}_2$ factors all have mass $m$.
However, the situation in this case is in detail different, with a variety of arguments suggesting individual upper bounds on both $m$ and $\Lambda$, of the form \cite{Dvali:2007hz,Dvali:2007wp} 
\begin{equation}
	m, \Lambda \lesssim \frac{M_{Pl}}{\sqrt{N}} \ .
\label{eq:DvaliZ2}
\end{equation}

The results of \cite{Dvali:2007hz,Dvali:2007wp,Dvali:2008tq} apply in the context of theories with discrete gauge symmetries at low energies, regardless of whether they are further UV-completed into an Abelian Higgs model, and are suggestive that some version of the WGC applies to symmetries that are discrete. We take a step towards making this expectation more precise by studying scenarios where the discrete group is realised through the spontaneous breaking of a $U(1)$ gauge symmetry. This leads us to the following observations in Abelian Higgs models that realize discrete symmetries in the infrared:
\begin{itemize}
\item Basic scaling properties of the effect of discrete hair on black holes \cite{Coleman:1991ku} reproduce Eq.(\ref{eq:Dvali}) for models of spontaneous symmetry breaking that feature either $\mathbb{Z}_N$ or $\mathbb{Z}_2^N$ gauge groups at low energies, with $\Lambda$ identified with the Higgs vev $v$.
In this context, a non-zero expectation value of the electric field is generated outside the event horizon due to a process of cosmic string loop nucleation on the surface of the BH \cite{Coleman:1991ku}. For small BHs, the effect of such virtual process is potentially unsuppressed, and could therefore provide a mechanism for discrete hair to be lost without requiring modifications in the gravitational sector. 

\item The infrared regime of an Abelian Higgs model admits a dual description in terms of two $U(1)$ gauge groups coupled through a topological term, with the additional $U(1)$ factor emerging out of a duality transformation \cite{Coleman:1991ku,Maldacena:2001ss,Banks:2010zn}. We show that applying the WGC to the dual picture one obtains individual upper bounds of both $m$ and $\Lambda \sim v$ that allow us to saturate but not violate Eq.(\ref{eq:Dvali}), for both $\mathbb{Z}_N$ or $\mathbb{Z}_2^N$ discrete groups. Moreover, in theories where the gauge group is $\mathbb{Z}_2^N$, considering the multi-field version of the conjecture \cite{Cheung:2014vva} further allows us to recover Eq.(\ref{eq:DvaliZ2}).
\end{itemize}

The dual picture of the Abelian Higgs model illuminates key features of the bounds in Eqs.(\ref{eq:Dvali})-(\ref{eq:DvaliZ2}). In particular, electric charge under the emergent $U(1)$ factor corresponds to the units of magnetic flux carried by cosmic strings in the Higgs description, and the appearance of both $m$ and $\Lambda \sim v$ in Eqs.(\ref{eq:Dvali})-(\ref{eq:DvaliZ2}) admits a simple physical interpretation: Light charged particles must be part of the spectrum for BHs to decay, whereas an upper bound on the tension of cosmic strings, and therefore on the scale of spontaneous symmetry breaking, ensures that there exists a mechanism through which discrete charge can be lost, namely the non-zero electric field that is generated when a virtual cosmic string loop wraps around the horizon of the BH.

It is worth emphasizing that Eqs.(\ref{eq:Dvali})-(\ref{eq:DvaliZ2}) only imply a bound on $m$ and/or $\Lambda$ parametrically below the Planck scale for $N \gg 1$. In the $\mathbb{Z}_2^N$ case this is readily understood, as the gravitational cutoff is reduced to $M_{Pl} / \sqrt{N}$ in theories containing $N$ stable species \cite{Dvali:2007wp}.
For a theory with a single $\mathbb{Z}_N$ gauge group it can be understood as an obstruction to recovering a global $U(1)$ in the large $N$ limit: In principle, one could take the limit $N \rightarrow \infty$ while keeping the vector mass $m_\gamma = g N v$ fixed, but this would require taking $g \propto 1 / N \rightarrow 0$. \footnote{In the context of a UV-completion into an Abelian Higgs model the Higgs charge is $gN$, so the global limit could a priori be taken while retaining perturbativity.} We would then recover a global $U(1)$, a limit we should not be able to take with impunity, as ensured through Eq.(\ref{eq:Dvali}).

The rest of this article is organized as follows: In section \ref{sec:remnants} we review the arguments of \cite{Dvali:2007hz,Dvali:2007wp,Dvali:2008tq} leading to Eqs.(\ref{eq:Dvali})-(\ref{eq:DvaliZ2}), and show how similar upper bounds can be obtained by considering the effect of discrete hair on BHs when the discrete symmetry is realised through Higgsing.
In section~\ref{sec:WGCdiscrete}, we review how a discrete gauge symmetry written in terms of an Abelian Higgs model is dual to a theory containing two $U(1)$ factors, and show how applying the WGC to the dual theory leads to non-trivial constraints consistent with those of section \ref{sec:remnants}.
Section~\ref{sec:naturalness} addresses ways in which the WGC applied to a $\mathbb{Z}_N$ gauge theory can lead to novel constraints on effective field theory parameters, with particular relevance to the apparent fine-tuning of the weak scale.
We present our conclusions in section~\ref{sec:conclusions}.

\section{No remnants stabilized by discrete charge}
\label{sec:remnants}

\subsection{Non-perturbative black hole arguments}
\label{sec:BHargs}

We now review the argument presented in \cite{Dvali:2007hz,Dvali:2008tq} leading to the upper bound in Eq.(\ref{eq:Dvali}) for theories with gauge group $\mathbb{Z}_N$.
In order to simplify the discussion, we assume that the only degree of freedom charged under the discrete symmetry is a state $\psi$, with mass $m$, which carries the unit of $\mathbb{Z}_N$ charge.
Below some cutoff scale, the effective description of the theory is that of $\psi$ particles with $\mathbb{Z}_N$-preserving interactions.

We start by considering a BH carrying $\sim N$ units of discrete charge, with initial size much larger than the Compton wavelength of the unit-charge particle. This ensures that the initial temperature of the BH is $T \ll m$, so that as the BH evaporates it does so without losing any of its charge (Hawking evaporation into particles with mass larger than the BH temperature is exponentially suppressed).
However, if we want there to be no Planckian objects stabilized by discrete hair at the end of the evaporation process, the BH must be able to shed its charge when it reaches some size $\Lambda^{-1}$, with $\Lambda$ an energy scale which a priori could be anywhere in the range $m \lesssim \Lambda \lesssim M_{Pl}$. At this stage, the mass of the BH is $M \sim M_{Pl}^2 / \Lambda$, which must be larger than $N$ times the mass of the unit-charge quantum.
By imposing this basic kinematic requirement one recovers Eq.(\ref{eq:Dvali}), i.e.~
$$ M \sim \frac{M_{Pl}^2}{\Lambda} \gtrsim N \cdot m \qquad \qquad \Rightarrow \qquad \qquad m \cdot \Lambda \lesssim \frac{M_{Pl}^2}{N} \ .$$

The physical effect of discrete electric charge on BH properties, first studied in \cite{Coleman:1991ku}, is exponentially suppressed in the regime where the theory can be described in terms of effective $\mathbb{Z}_N$-preserving interactions among light degrees of freedom. So long as this effective description is valid, the behaviour of a BH carrying discrete charge is approximately the same as for the Schwarzschild solution, as noted in \cite{Dvali:2007hz,Dvali:2008tq}. In particular, up to exponentially small corrections, no electric field exists outside the BH horizon, and Hawking evaporation proceeds equally into $\psi$ and $\bar \psi$ quanta. As a result, in this regime, the BH will not be able to shed its discrete hair. The scale $\Lambda$ must therefore be identified with a cutoff scale beyond which such an effective description is no longer valid \cite{Dvali:2007hz,Dvali:2008tq}. Specifically, new effects must become relevant at the scale $\Lambda$ that allow the BH to shed its discrete hair, significantly deviating from its previous Schwarzschild behaviour. In \cite{Dvali:2008tq} a stronger constraint on $\Lambda$ is advocated for by making the additional assumption that the hair loss mechanism is intrinsically thermal, and that a BH of size $R \sim \Lambda^{-1}$ radiates preferentially into modes with energies of order $\Lambda$. This leads to the bound $\Lambda \lesssim M_{Pl} / \sqrt{N}$, which with the implicit constraint $m \lesssim \Lambda$ combines with Eq.(\ref{eq:Dvali}) to give an identical bound on $m$. Although this assumption may hold in other realizations of a discrete gauge symmetry, it appears unjustified in the context of an Abelian Higgs model, as we discuss in section~\ref{sec:discretehair}. We therefore regard Eq.(\ref{eq:Dvali}) as the only solid upper bound applicable to $\mathbb{Z}_N$ gauge theories.

The same general argument applies when the gauge group is $\mathbb{Z}_2^N$ instead of $\mathbb{Z}_N$, leading to the same upper bound as in Eq.(\ref{eq:Dvali}), under the assumption that all the states $\psi_i$ carrying the unit of charge under the $N$ individual $\mathbb{Z}_2$ factors have the same mass $m$ \cite{Dvali:2007hz,Dvali:2007wp}.
In this case, the argument can be taken a step further by making the observation that in theories containing $N$ stable species the gravitational cutoff is expected to be lowered down to $M_{Pl} / \sqrt{N}$ \cite{Dvali:2007wp}. The requirement that both $m$ and $\Lambda$ are below the gravitational cutoff of the theory therefore leads to individual upper bounds of the form
$$ m, \Lambda \lesssim \frac{M_{Pl}}{\sqrt{N}} \ ,$$
as in Eq.(\ref{eq:DvaliZ2}). \footnote{The upper bound on $m$ can also be partially justified by noting that the natural value of the Planck scale in a theory containing $N$ particles with mass $m$ is expected to satisfy $M_{Pl}^2 \gtrsim N \cdot m^2$ \cite{Adler:1980pg,Zee:1981mk,Dvali:2007hz}.}

The authors of \cite{Dvali:2007hz,Dvali:2007wp,Dvali:2008tq} further speculate that in both cases the UV completion required at the scale $\Lambda$ must be necessarily linked to the gravitational sector.
Although there might be realisations of a $\mathbb{Z}_N$ gauge symmetry where this is the case, if the theory is implemented through an Abelian Higgs model, a more plausible expectation is that $\Lambda$ may just correspond to the scale of spontaneous symmetry breaking, as we discuss next. 

\subsection{Effect of discrete hair on black holes}
\label{sec:discretehair}

The effect of electric  $\mathbb{Z}_N$ charge on BH thermodynamics was first explored by Coleman, Preskill, and Wilczek in \cite{Coleman:1991ku}, where the discrete gauge symmetry is realised through a model of spontaneous symmetry breaking.
It is argued in \cite{Coleman:1991ku} that the leading effect of discrete hair on BHs is due to a process whereby a cosmic string loop is nucleated on the horizon surface, grows to envelope the BH, and re-annihilates at the antipodal point. In the Euclidean path integral approach to BH thermodynamics, this process is expressed through the existence of instantons, specifically vortex solutions on the $r - \tau$ plane ($\tau$ being the periodic Euclidean time coordinate). Different units of magnetic flux wrapping the BH in the space-time picture correspond to vortices with different winding numbers of the scalar field in the Euclidean formalism.

When computing the partition function relevant for a description of the system in the context of a given thermodynamic ensemble, one must therefore also sum over sectors with different scalar field vorticities.
In the canonical ensemble, where both temperature $\beta^{-1}$ and charge $Q$ are fixed, the partition function factors into a discrete sum of the form \cite{Coleman:1991ku}
\begin{equation}
	Z(\beta, Q) \sim \sum_{k=-\infty}^{+ \infty} e^{i 2 \pi Q k / N} Z_k (\beta, Q) \ ,
\label{eq:ZQ}
\end{equation} 
where $Z_k$ represents the partition function in a sector with winding number $k$, and appears in the sum weighted by a phase factor $e^{i 2 \pi Q k / N}$, corresponding to the Aharonov-Bohm phase that a string carrying $k$ units of flux picks up as it envelops a BH with charge $Q$.

Field configurations with higher winding number have higher action costs, and, as a result, Eq.(\ref{eq:ZQ}) can be evaluated in a semiclassical expansion. The sum is dominated by the $k=0$ configuration, and contributions from the sectors with $k = \pm 1$ provide the leading charge-dependent correction. To leading order, the effect of $\mathbb{Z}_N$ charge on the thermodynamic partition function can be written as, parametrically \cite{Coleman:1991ku}
\begin{equation}
	\log \left( \frac{Z (\beta, Q)}{Z(\beta)} \right)^{-1} \sim \left[ 1 - \cos \left( \frac{2 \pi Q}{N} \right) \right] e^{-\Delta S_v} \ ,
\label{eq:Zdiscrete}
\end{equation}
where $\Delta S_v$ represents the action of the $k=1$ vortex, and the correction vanishes whenever $Q$ is an integer multiple of $N$, as expected.

Of particular interest is the status of the electric field generated by a BH carrying discrete hair. At the classical level, i.e.~for $k=0$, the electric field vanishes.
At the quantum level this is no longer true, and a non-zero expectation value of the electric field operator in the radial direction arises when taking into account the contribution from sectors of non-zero winding number.
As before, this can be computed in a semiclassical expansion, and, to leading order, one obtains an expression of the form \cite{Coleman:1991ku}:
\begin{equation}
	\langle E_r (r) \rangle \sim \sin \left( \frac{2 \pi Q}{N} \right) F_{r \tau} (r) e^{-\Delta S_v} \ ,
\label{eq:Er}
\end{equation}
where $F_{r \tau} (r)$ corresponds to the Euclidean magnetic field component for the unit-flux vortex.
This expression is consistent with the space-time picture put forward in \cite{Coleman:1991ku}: as the string wraps around the BH, the moving magnetic flux inside the string generates an electric field in the radial direction.
This electric field is only non-zero when a string is nucleated, and therefore its expectation value is suppressed by a factor related to the corresponding tunnelling rate, with such suppression provided by the last factor in Eq.(\ref{eq:Er}).

The validity of the semiclassical expansion remains so long as $\Delta S_v \gg 1$, and there are two limiting regimes in which an analytic expression for the vortex action can be obtained: the thin- and thick-string limits, corresponding to the thickness of the string being much smaller or much larger than the size of the BH.
\footnote{Both limits were discussed in \cite{Coleman:1991ku}, with \cite{Dowker:1991qe} and \cite{GarciaGarcia:2018tua} providing important extra considerations in the thin- and thick-string regimes respectively.}
These two regimes are realised, respectively, for BH sizes much bigger or much smaller than the Compton wavelength of the vector.
In particular, the vortex action in the thin-string limit is given by, parametrically
\begin{equation}
	\Delta S_v \sim r_+^2 v^2 \ ,
\end{equation}
with $r_+$ the horizon radius. This expression intuitively corresponds to the tension of the string $T_s \sim v^2$ times the corresponding world-sheet area.
This is the regime of interest for large BHs, with sizes $r_+ \gg v^{-1}$, in which case the discrete symmetry is well approximated through an effective description in terms of light states with $\mathbb{Z}_N$-preserving interactions. As anticipated in section \ref{sec:BHargs}, the properties of a BH carrying discrete charge are therefore much like those of its uncharged counterpart, up to exponentially small corrections.

However, the semiclassical expansion will break down as the BH gets small.
If the separation of scales between the vector mass and the scale of spontaneous symmetry breaking is not too large, i.e.~if the coupling $gN$ is not much smaller than 1, then the semiclassical analysis breaks down as we approach the thick-string limit, and the size of the BH becomes $r_+ \sim v^{-1}$.
In this regime, the effect of discrete hair on such small BHs is potentially large. In particular, the existence of an unsuppressed electric field would allow the BH to lose its hair by the same mechanism through which the Reissner-N\"ordstrom solution sheds its charge, i.e.~through a combination of Schwinger pair production of particle and anti-particle pairs, and asymmetric Hawking evaporation.
Demanding that a BH of size $v^{-1}$, and therefore mass $M \sim M_{Pl}^2 / v$, is kinematically allowed to lose $\sim N$ units of discrete hair requires
\begin{equation}
	m \cdot v \lesssim \frac{M_{Pl}^2}{N} \ ,
\label{eq:discretehair}
\end{equation}
which coincides with Eq.(\ref{eq:Dvali}) after identifying $\Lambda \sim v$.

We expect Eq.(\ref{eq:discretehair}) to also approximately hold when the discrete group is $\mathbb{Z}_2^N$, instead of $\mathbb{Z}_N$.
In this case, the appropriate version of Eq.(\ref{eq:Zdiscrete}) now reads
\begin{equation}
	\log \left( \frac{Z (\beta, Q)}{Z(\beta)} \right)^{-1} \sim N \left[ 1 - \cos \left( \pi Q \right) \right] e^{-\Delta S_v} \ ,
\end{equation}
and the electric field of Eq.(\ref{eq:Er}) would now include $N$ different components, corresponding to the individual $\mathbb{Z}_2$ factors.
Despite the $N$-dependent enhancement present in this case, the expression for $\Delta S_v$ is the same as in the $\mathbb{Z}_N$ case, assuming all $\mathbb{Z}_2$ factors are UV-completed into an Abelian Higgs model at the same scale $v$.
Thus, up to corrections scaling as $\log N$, we obtain the same parametric bound as in Eq.(\ref{eq:discretehair}).

We emphasize that verifying the expectation that small BHs are capable of losing their hair as a result of cosmic string loop nucleation would require a dedicated calculation well beyond the regime in which existing results are applicable \cite{Coleman:1991ku,Dowker:1991qe,GarciaGarcia:2018tua}.
Our discussion, however, raises the more conservative possibility that $\Lambda$, understood as a proxy for the scale at which physics responsible for discrete hair loss must appear, may simply correspond to the scale of UV completion into an Abelian Higgs model, as opposed to being intrinsically related to physics of the gravitational sector.

\section{Towards a Weak Gravity Conjecture for discrete gauge symmetries}
\label{sec:WGCdiscrete}

We now explore the extent to which the above bounds on discrete gauge symmetries may be further understood by applying the WGC in the context of Abelian Higgs models and their dual descriptions. In section~\ref{sec:duality} we review how a discrete gauge theory realised in terms of a spontaneously broken $U(1)$ admits a dual description in terms of two $U(1)$ factors coupled through a topological term.
In~\ref{sec:conjecture} we show how applying the electric form of the WGC in the dual picture leads to constraints consistent with those of Eq.(\ref{eq:Dvali}) and (\ref{eq:DvaliZ2}), so long as $\Lambda$ is identified with the scale of spontaneous symmetry breaking.

\subsection{Dual description of Higgsing}
\label{sec:duality}

An equivalent description of a $\mathbb{Z}_N$ gauge theory at low energies can be written in terms of two $U(1)$ gauge groups coupled through a topological term \cite{Coleman:1991ku,Maldacena:2001ss,Banks:2010zn}. We refer to the two Abelian factors as $U(1)_A$ and $U(1)_B$, with the associated gauge potentials being a 1-form $A$ and a 2-form $B$.
The gauge couplings corresponding to the $U(1)_A$ and $U(1)_B$ factors are $g$ and $f$, and have mass dimensions 0 and 1 respectively.

The effective lagrangian contains kinetic terms for both gauge fields, as well as a $B \wedge F$ coupling of the form \cite{Banks:2010zn}
\begin{equation}
	\mathcal{L} \supset - \frac{1}{12 f^2} H_{\mu \nu \rho} H^{\mu \nu \rho} - \frac{1}{4 g^2} F_{\mu \nu} F^{\mu \nu}  - \frac{N}{8 \pi} \varepsilon^{\mu \nu \rho \sigma} B_{\mu \nu} F_{\rho \sigma} \ ,
\label{eq:lagIR}
\end{equation}
where $F \equiv dA$, $H \equiv dB$, and we have used a normalization where the gauge coupling sits in front of the kinetic term.
In this description, the order of the discrete symmetry corresponds to the integer $N$ in front of the $B \wedge F$ coupling.
The lagrangian may contain further interaction terms, not included in Eq.(\ref{eq:lagIR}), involving charged dynamical objects, with electric charge under $U(1)_A$ and $U(1)_B$ quantized in units of $g$ and $f$ respectively.

Objects carrying electric $U(1)_A$ charge correspond to point particles, whereas those charged under $U(1)_B$ are strings.
Wilson loop operators describing the dynamics of charged objects are given by
\begin{equation}
	W_A (q, C) = \exp \left( i q \oint_C A \right) \ , \qquad {\rm and} \qquad W_B (k, \Sigma) = \exp \left( i k \oint_\Sigma B \right) \ ,
\label{eq:WilsonLoops}
\end{equation}
for a particle with world-line $C$ carrying $q$ units of $U(1)_A$ charge, and a string with world-sheet $\Sigma$ carrying $k$ units of $U(1)_B$ charge, respectively.

Although the operators in Eq.(\ref{eq:WilsonLoops}) can be defined for any $q, k \in \mathbb{Z}$, only operators defined modulo $N$ label different observables.
In physical terms, the only long-range interaction available to measure $U(1)_A$ and $U(1)_B$ charge is Aharonov-Bohm scattering of particles and strings.
The Aharonov-Bohm phase that a string carrying $k$ units of electric $U(1)_B$ charge picks up when looped around a particle carrying $U(1)_A$ charge $q$ is given by
\begin{equation}
	\exp \left( i \frac{2 \pi q k}{N} \right) \ .
\label{eq:ABphase}
\end{equation}
Eq.(\ref{eq:ABphase}) makes it manifest that observables in this theory are labelled in $\mathbb{Z}_N$, with electric $U(1)_A$ and $U(1)_B$ charge only defined modulo $N$.
Indeed, a gauge invariant order parameter that would distinguish the $\mathbb{Z}_N$ phase of the theory from the completely broken one (i.e.~$N=1$) necessarily involves the two types of operators
in Eq.(\ref{eq:WilsonLoops}), with $C$ intersecting $\Sigma$ once \cite{Preskill:1990bm}, essentially capturing the operation described above.
\footnote{We note that to define an order parameter distinguishing the broken and unbroken phases of a $\mathbb{Z}_N$ gauge symmetry it is only technically necessary to make reference to probe particles and strings -- for this purpose, charged objects need not be dynamical.}

The theory of Eq.(\ref{eq:lagIR}) is dual to a more familiar realization of a $\mathbb{Z}_N$ gauge symmetry: an Abelian Higgs model where the charge of the condensate is $N$ times the charge quantum.
To make this manifest it is convenient to rewrite the last term in Eq.(\ref{eq:lagIR}) in terms of $H$ (after integration by parts), and introduce an extra degree of freedom $\varphi$ acting as a Lagrange multiplier that enforces the exactness of $H$ dynamically. After rescaling the gauge fields so that their kinetic terms are canonically normalized, the lagrangian reads
\begin{equation}
	\mathcal{L} \supset - \frac{1}{12} H_{\mu \nu \rho} H^{\mu \nu \rho} - \frac{1}{4} F_{\mu \nu} F^{\mu \nu} + \frac{Ngf}{12 \pi} \varepsilon^{\mu \nu \rho \sigma} H_{\mu \nu \rho} A_{\sigma} 
				- \frac{1}{6} \varphi \varepsilon^{\mu \nu \rho \sigma} \partial_\mu H_{\nu \rho \sigma} \ . 
\label{eq:lagDual}
\end{equation}
We may now integrate out $H$ by using the corresponding equations of motion, which read
\begin{equation}
	H^{\mu \nu \rho} = - \varepsilon^{\mu \nu \rho \sigma} (\partial_\sigma \varphi - \frac{N g f}{2 \pi} A_\sigma) \ .
\label{eq:Heom}
\end{equation}
Finally, plugging Eq.(\ref{eq:Heom}) back into (\ref{eq:lagDual}), one finds
\begin{equation}
	\mathcal{L} \supset - \frac{1}{2} (\partial_\mu \varphi - N g v A_\mu)^2 - \frac{1}{4} F_{\mu \nu} F^{\mu \nu} \ ,
\label{eq:Higgs}
\end{equation}
where we have defined $v \equiv f / 2 \pi$.

Eq.(\ref{eq:Higgs}) is precisely the effective lagrangian describing the spontaneous breaking of a continuous Abelian gauge theory through a Higgs field with vev $v$, after integrating out the radial mode.
\footnote{Strictly speaking, the far infrared regime of a discrete gauge symmetry admits a universal description as given by the last term in Eq.(\ref{eq:lagIR}), corresponding to a purely topological theory \cite{Banks:2010zn}.
We note that such a theory admits UV completions different than an Abelian Higgs model, although we will not consider any specific alternatives in this work.}
The energy scale defined by the gauge coupling of the $U(1)_B$ factor corresponds in the Abelian Higgs model to the scale of spontaneous symmetry breaking. In the dual picture, strings carrying $k$ units of electric $U(1)_B$ charge correspond in the Higgs description to cosmic strings threaded by $k$ units of magnetic flux.

Our discussion makes it manifest that electric $U(1)_A$ charge matches on to charge under the spontaneously broken $U(1)$ in the Abelian Higgs model. This is the type of charge we have referred to as discrete hair, which BHs can carry \cite{Krauss:1988zc,Coleman:1991ku}. It is then natural to wonder whether BHs may also carry electric $U(1)_B$ hair.
Indeed, non-trivial solutions to the Einstein-Maxwell-Higgs equations exist that correspond to a BH threaded by an infinitely long flux-tube \cite{Aryal:1986sz,Achucarro:1995nu},
the $U(1)_B$ charge of the configuration corresponding to the units of flux confined inside the string.
Although there is a sense in which such charge may be regarded as BH hair \cite{Achucarro:1995nu}, it is clearly different from the type of hair we have so far been discussing, and  it is unclear how arguments similar to those of \cite{Dvali:2007hz,Dvali:2007wp,Dvali:2008tq} could be made in this context.
We will not discuss this type of solutions further in this work, but note that a more detailed consideration of their properties might provide valuable intuition.
\footnote{Application of the arguments discussed here in the context of higher-dimensional black objects, such as black strings, carrying $U(1)_B$ hair may provide further insight.}

If the gauge group is $\mathbb{Z}_2^N$, instead of $\mathbb{Z}_N$, the equivalent version of Eq.(\ref{eq:lagIR}) is given by
\begin{equation}
	\mathcal{L} \supset - \sum_{j=1}^N \left( \frac{1}{12 f^2} H_{j \mu \nu \rho} H_j^{\mu \nu \rho} + \frac{1}{4 g^2} F_{j \mu \nu} F_j^{\mu \nu}  + \frac{1}{4 \pi} \varepsilon^{\mu \nu \rho \sigma} B_{j \mu \nu} F_{j \rho \sigma} \right) \ ,
\label{eq:lagZ2N}
\end{equation}
where, for simplicity, we have assumed that the gauge couplings of the $N$ $U(1)_A$ and $U(1)_B$ factors are all equal to $g$ and $f$ respectively.
Eq.(\ref{eq:lagZ2N}) is dual to $N$ copies of a $U(1)$ gauge symmetry spontaneously broken down to $\mathbb{Z}_2$ by Higgs fields with vev $v = f / 2 \pi$, and charge twice the electric charge quantum.

\subsection{Weak Gravity Conjecture}
\label{sec:conjecture}

The value of rewriting the effective description of a discrete gauge theory in terms of the dual picture described in section~\ref{sec:duality} lies on the fact that electric $\mathbb{Z}_N$ charge and magnetic flux quanta are put on the same footing: in the dual description, they correspond to electric charge of the $U(1)_A$ and $U(1)_B$ factors, respectively.
Although $U(1)_A$ and $U(1)_B$ charge is only defined modulo $N$, and only their product can be measured at long distances, the dual description suggests that if there is a version of the WGC that applies for theories with discrete gauge groups, it might involve both $U(1)$ factors.

We can gain some insight into how a version of the WGC for theories with $\mathbb{Z}_N$ gauge group may ultimately arise by applying the electric form of the conjecture in the dual picture to both $U(1)_A$ and $U(1)_B$.\footnote{Of course, in the dual picture with gauge potentials $A$ and $B$, we can further dualize $A$ to another 1-form that may be interpreted as a matter field Higgsing the $U(1)_B$ symmetry down to $\mathbb{Z}_N$ \cite{Banks:2010zn}. In this case one may question the utility of applying the WGC to $U(1)_A$ and $U(1)_B$, insofar as there is always another duality frame in which one is a matter field responsible for Higgsing the other. There are, however, a variety of arguments in favor of applying the WGC to this scenario (see e.g.~\cite{Reece:2018zvv}), and we proceed apace.}
Doing so translates into the requirement that there be particles with mass $m$ charged under $U(1)_A$, and strings with tension $T_s$ charged under $U(1)_B$ satisfying
\begin{equation}
	m \lesssim g M_{Pl} \ , \qquad \qquad {\rm and} \qquad \qquad T_s \lesssim f M_{Pl} \ .
\label{eq:WGCv1}
\end{equation}

We can further interpret Eq.(\ref{eq:WGCv1}) in the context of a spontaneously broken $U(1)$ gauge theory. In this case, $f \sim v$ and $T_s \sim v^2$, and the upper bound on $T_s$ translates into
\begin{equation}
	v \lesssim M_{Pl} \ , 
\label{eq:vBound}
\end{equation}
i.e.~the requirement that the theory be UV-completed into an Abelian Higgs model at energies at or below the gravitational cutoff.
Moreover, for the theory to remain perturbative we must have $g \lesssim 1 /N $, and so the upper bound on $m$ in turn implies the (weaker) constraint
\begin{equation}
	m \lesssim \frac{M_{Pl}}{N} \ .
\label{eq:mBound}
\end{equation}
Combining Eqs.(\ref{eq:vBound})-(\ref{eq:mBound}) it follows that
$$
	m \cdot v \lesssim \frac{M_{Pl}^2}{N} \ ,
$$
which coincides with Eq.(\ref{eq:discretehair}).

In a weakly coupled theory, the upper bound on $m$ in Eq.(\ref{eq:WGCv1}) is more stringent than that of Eq.(\ref{eq:mBound}), and this, in turn, stronger than that implied by the upper bound on the product of $m \cdot v$ as given in Eq.(\ref{eq:discretehair}) (indeed, Eq.(\ref{eq:discretehair}) allows $m$ to be as high as $M_{Pl} / \sqrt{N}$ so long as the theory is UV-completed into an Abelian Higgs model at roughly the same scale).
The crucial insight, however, is that by applying the WGC to both the $U(1)_A$ and $U(1)_B$ factors we obtain constraints on the mass and tension of charged objects that allow us to \emph{saturate but not violate} the upper bound Eq.(\ref{eq:discretehair}) based on BH decay arguments.

A more sophisticated treatment might lead to Eq.(\ref{eq:discretehair}) as the `true' version of the WGC that applies to theories with $\mathbb{Z}_N$ gauge group -- perhaps an argument could be made that the WGC bound should only be applied to the product of $m$ and $T_s$, as opposed to individually as in Eq.(\ref{eq:WGCv1}), based on the observation that only the product of $U(1)_A$ and $U(1)_B$ charge corresponds to an asymptotic observable.

Discussing the WGC in the context of the dual theory also clarifies some of the differences between $\mathbb{Z}_N$ and $\mathbb{Z}_2^N$ theories.
In the latter case, applying the conjecture to a theory with $N$ copies of the $U(1)_A$ and $U(1)_B$ factors leads to
\begin{equation}
	m \lesssim \frac{g M_{Pl}}{\sqrt{N}} \ , \qquad \qquad {\rm and} \qquad \qquad T_s \lesssim \frac{f M_{Pl}}{\sqrt{N}}\ ,
\label{eq:WGCmultiv1}
\end{equation}
where the factors of $\sqrt{N}$ correspond to the modification of the WGC when applied to theories with multiple gauge groups \cite{Cheung:2014vva}, and we have made the simplifying assumption that all unit-charge particles and strings have the same mass $m$, and tension $T_s$ respectively.
In an Abelian Higgs model, $f \sim v$ and $T_s \sim v^2$ as before, but now the perturbativity requirement $g \lesssim 1/2 \sim 1$ in independent of $N$. Eq.(\ref{eq:WGCmultiv1}) therefore implies independent upper bounds on both $m$ and $v$, of the form
\begin{equation}
	m , v \lesssim \frac{M_{Pl}}{\sqrt{N}} \ ,
\label{eq:WGCmultiv2}
\end{equation}
which coincides with Eq.(\ref{eq:DvaliZ2}) so long as $\Lambda$ is identified with the Higgs vev. This result further fits in well with the expectation that the gravitational cutoff is lowered down to $M_{Pl} / \sqrt{N}$ for a $\mathbb{Z}_2^N$ gauge theory. In this context, Eq.(\ref{eq:WGCmultiv2}) just corresponds to the reasonable statement that both the mass of the unit-charge particle, and the scale of spontaneous symmetry breaking, must be below the gravitational cutoff. 


\section{Naturalness in the Swampland}
\label{sec:naturalness}

The aim behind the Swampland Program is to identify consistency conditions that effective field theories need to satisfy in order to be compatible with a further UV completion into a theory of quantum gravity \cite{Vafa:2005ui}.
Much of the intuition behind Swampland criteria, most of which remain in conjectural form, stems from patterns observed in known string theory constructions, with infrared arguments based on BH physics, when available, providing extra support (a list of current Swampland conjectures includes \cite{Vafa:2005ui,Ooguri:2006in,ArkaniHamed:2006dz,Adams:2006sv,Ooguri:2016pdq,Montero:2017yja,Danielsson:2018ztv,Obied:2018sgi}).
The WGC is perhaps the best-known condition in the list of Swampland criteria, and potential implications of the conjecture have been studied in a variety of circumstances.
For example, variations of the conjecture can lead to significant constraints on models of inflation that require super-Planckian field ranges \cite{delaFuente:2014aca,Brown:2015iha}, rule out models that include parametrically light St\"uckelberg massive photons \cite{Reece:2018zvv}, or even explain the apparent fine-tuning of the weak scale through arguments linking it to the size of the cosmological constant \cite{Ibanez:2017kvh,Ibanez:2017oqr}.
Although these statements crucially depend on additional assumptions, the most obvious being the specific form of the conjecture to be applied in the low energy regime of effective theories
\cite{Saraswat:2016eaz,Choi:2015fiu,Kaplan:2015fuy,Craig:2018yld}, such efforts are valuable to the extent that they link our intuition about quantum gravity to experimental observation.
\footnote{Low energy implications stemming from alternative versions of the conjecture, such as the scalar WGC proposed in \cite{Palti:2017elp}, have also been identified and discussed in \cite{Lust:2017wrl}.}

Indeed, the form of Eq.(\ref{eq:WGC}), which sets an upper bound on the mass of a state charged under an Abelian gauge symmetry, raises the question of whether the conjecture could provide an explanation of the smallness of the weak scale compared to $M_{Pl}$.
Although not directly applicable to the SM Higgs, it could still be applied to a different state whose mass arises from the Higgs vev, so that Eq.(\ref{eq:WGC}) would translate into an indirect constraint on $v_{\rm SM}$.
The simplest version of this idea, proposed in \cite{Cheung:2014vva}, is realised if the $U(1)_{B-L}$ symmetry of the SM is gauged, and the constraint Eq.(\ref{eq:WGC}) applied to one of the neutrinos, i.e.~$m_\nu \lesssim g_{B-L} M_{Pl}$. Since $m_\nu \lesssim 0.1 \ {\rm eV}$ this would require $g_{B-L} \lesssim 10^{-28}$, which is consistent with the current experimental upper bound of $10^{-24}$ \cite{Adelberger:2009zz,Wagner:2012ui}.
If neutrino masses arise through a Yukawa coupling to the SM Higgs, then the weak scale would be indirectly constrained by the WGC.

The above argument employing the electric WGC is, however, undone by its magnetic counterpart. Specifically, the magnetic version of the conjecture, which applies to unbroken $U(1)$ gauge theories, demands some form of UV completion at the scale $g M_{Pl}$, which coincides with the scale at which the particle satisfying the WGC is present if the upper bound on its mass is saturated.
Obvious ways of addressing this issue, such as embedding the $U(1)_{B-L}$ into an $SU(2)$ gauge group at the scale in question, are clearly in conflict with experimental observation.

This difficulty disappears if $U(1)_{B-L}$ is broken down to a gauge $\mathbb{Z}_N$ subgroup.
\footnote{This possibility was first suggested in \cite{Dvali:2008tq}, but as we will discuss momentarily, the different conjectural bounds on $m$ and $\Lambda$ in \cite{Dvali:2008tq} lead to different conclusions.} Now, the scale at which particles carrying $\mathbb{Z}_N$ charge appear may be parametrically below the scale of UV completion into an Abelian Higgs model. If $U(1)_{B-L}$ is spontaneously broken at a scale $v \sim M_{Pl}$, 
then
electrically charged states must be present below $\sim M_{Pl} / N$.
As applied to neutrinos, this leads to
\begin{equation}
	m_\nu \lesssim \frac{M_{Pl}}{N} \ ,
\label{eq:mnu}
\end{equation}
which with $N \gtrsim 10^{28}$ requires $m_\nu \lesssim 0.1$ eV.
If the neutrino mass arises through a tiny Yukawa coupling with the SM Higgs, $m_\nu \sim y_\nu v_{\rm SM}$, then Eq.(\ref{eq:mnu}) implies
\begin{equation}
	v_{\rm SM} \lesssim \frac{M_{Pl}}{y_\nu N} \ ,
\label{eq:vsm}
\end{equation}
and the correct value of the weak scale is obtained for $y_\nu \sim 10^{-12}$.

Notice that even though Eq.(\ref{eq:vsm}) is bounding the observed value of the weak scale by parameters $N$ and $y_\nu$ that are technically natural, the theory would still appear fine-tuned from an effective field theory perspective, since the cutoff scale of the theory, $M_{Pl}$, is well above $v_{\rm SM}$.
This is therefore not so much a \emph{solution} to the hierarchy problem, but rather an \emph{explanation} of why nature appears finely-tuned: those versions of the theory that are natural would fall into the Swampland of theories that do not satisfy the WGC as applied to discrete gauge symmetries.
Our suggestion therefore lacks an experimental `smoking-gun' signature that often models of naturalness provide. However, we note that the scenario proposed here would be falsified if (neutrinoless) double-beta decay was experimentally observed, as a Majorana mass for the neutrinos would break $U(1)_{B-L}$ down to a $\mathbb{Z}_2$ factor, therefore invalidating our argument.

A potentially more satisfying approach would involve lowering the gravitational cutoff,
therefore not only constraining the experimental value of the weak scale but also making the theory natural, in the traditional sense of lowering the scale of radiative corrections to the Higgs mass parameter.
This can be achieved if the theory contains $N$ copies of a $\mathbb{Z}_2$ symmetry, with $N \sim 10^{32}$, as first noted in \cite{Dvali:2007hz}.
In the context of a \emph{single} discrete symmetry such an approach was first suggested in \cite{Dvali:2008tq}, as applied to a $\mathbb{Z}_N$ $B-L$ gauge group using the the weaker bound $m_\nu \lesssim M_{Pl} / \sqrt{N}$ advocated therein. In this case, fixing the value of the weak scale via a bound on the neutrino mass would require $N \gtrsim 10^{56}$, with Eq.(\ref{eq:Dvali}) further implying that the theory requires a UV completion at roughly the same scale.
This suggestion has several difficulties. If the discrete symmetry is realised through an Abelian Higgs model, then $\Lambda \sim v$, and a scalar excitation carrying $\mathcal{O}(1)$ $U(1)_{B-L}$ charge would have to be present at roughly the scale of the neutrino mass. This is in conflict with experiment, and, as discussed in section \ref{sec:remnants}, such a UV completion does not require a gravitational cutoff at scales parametrically below $M_{Pl}$.
Even if one insisted on a UV completion in which $\Lambda$  corresponds to the scale at which gravitational effects become important, as advocated in \cite{Dvali:2008tq}, then a gravitational cutoff would be at the scale of the neutrino mass, obviously in contradiction with experimental observations.

Finally, we acknowledge that $\mathbb{Z}_N$ gauge symmetries with parametrically large $N$ may be hard to realise while maintaining an arbitrarily high cutoff scale. Attempts at finding controlled string constructions with parametrically large charges have not been successful \cite{Ibanez:2017vfl}. However, to the best of our knowledge, no consistency requirements of a theory of quantum gravity forbids a scenario with a Higgs field carrying charge $N \ggg 1$.

\section{Conclusions}
\label{sec:conclusions}

Arguments based on the absence of BH remnants stabilized by discrete charge lead to non-trivial bounds on discrete gauge symmetries \cite{Dvali:2007hz,Dvali:2007wp,Dvali:2008tq} that are reminiscent of the WGC. In this work we have pursued deeper connections between these two ideas in the context of Abelian Higgs models that realize discrete gauge symmetries in the infrared. We have shown that the bounds of \cite{Dvali:2007hz,Dvali:2007wp,Dvali:2008tq} are consistent with expectations based on the effect of discrete hair on BHs in theories with spontaneously broken symmetries. Moreover, we have shown that the results of \cite{Dvali:2007hz,Dvali:2007wp,Dvali:2008tq} follow from applying the WGC to a dual description of the Abelian Higgs model in terms of two $U(1)$ factors coupled through a topological term. Specifically, individual upper bounds on the mass and tension of charged objects obtained by demanding the WGC holds allow us to saturate but not violate the constraints of \cite{Dvali:2007hz,Dvali:2007wp,Dvali:2008tq} based on black hole arguments.
This highlights the sense in which conjectured bounds on discrete gauge symmetries may be thought of as residuals of the WGC that survive in the Higgsed phase of continuous gauge theories.

Applying the WGC to a dual description of the Abelian Higgs model suggests a novel way in which a $\mathbb{Z}_N$ $B-L$ gauge symmetry could help explain the apparent fine-tuning of the weak scale, if neutrinos saturate the version of the WGC bound applicable to discrete gauge symmetries, namely $m_\nu \lesssim M_{Pl} / N$ with $N \sim 10^{28}$. Although this requires an extremely large value of $N$ that may be hard to realise in full-fledged string constructions, the choice is nevertheless technically natural, and provides another example of how Swampland conjectures may impact of our understanding of physics at low energies.

Of course, discrete symmetries need not originate from spontaneously broken continuous ones, and extending the discussion presented here to other UV completions is bound to shed further light on the relationship between various conjectured bounds on continuous and discrete gauge symmetries. For example, although cosmic strings arise semi-classically in the context of Abelian Higgs models, this is not true of other realizations of a $\mathbb{Z}_N$ gauge symmetry. Our discussion, however, suggests that dynamical strings with finite tension \emph{must} must be part of the spectrum in any theory where a discrete gauge symmetry remains at low energies: Otherwise, the process of string loop nucleation required to generate a non-zero electric field outside the event horizon, thereby allowing the BH to lose its discrete hair, would not be available. This fits in well with the completeness hypothesis  \cite{Polchinski:2003bq}, which also applies to discrete gauge groups (indeed, the proof of \cite{Harlow:2018jwu} applies to both continuous and discrete symmetries). There is clearly a lot more to be understood regarding the role of the WGC in the context of theories where the gauge group is discrete.

\section*{Acknowledgments}
We are grateful to Daniel Harlow for valuable comments on earlier versions of this work.
The research of IGG is funded by the Gordon and Betty Moore Foundation through Grant GBMF7392. The research of NC and SK is supported in part by the US Department of Energy under the Early Career Award DE-SC0014129 and the Cottrell Scholar Program through the Research Corporation for Science Advancement. Research at KITP is supported in part by the National Science Foundation under Grant No.~NSF PHY-1748958.

\bibliography{WGCdiscrete_refs}

\newcommand{\noop}[1]{}
\providecommand{\href}[2]{#2}\begingroup\raggedright\begin{thebibliography}{10}

\bibitem{ArkaniHamed:2006dz}
N.~Arkani-Hamed, L.~Motl, A.~Nicolis and C.~Vafa, \emph{{The String landscape,
  black holes and gravity as the weakest force}},
  \href{http://dx.doi.org/10.1088/1126-6708/2007/06/060}{\emph{JHEP} {\bf 06}
  (2007) 060}, [\href{https://arxiv.org/abs/hep-th/0601001}{{\tt
  hep-th/0601001}}].

\bibitem{Abbott:1989jw}
L.~F. Abbott and M.~B. Wise, \emph{{Wormholes and Global Symmetries}},
  \href{http://dx.doi.org/10.1016/0550-3213(89)90503-8}{\emph{Nucl. Phys.} {\bf
  B325} (1989) 687--704}.

\bibitem{Coleman:1989zu}
S.~R. Coleman and K.-M. Lee, \emph{{WORMHOLES MADE WITHOUT MASSLESS MATTER
  FIELDS}}, \href{http://dx.doi.org/10.1016/0550-3213(90)90149-8}{\emph{Nucl.
  Phys.} {\bf B329} (1990) 387--409}.

\bibitem{Holman:1992us}
R.~Holman, S.~D.~H. Hsu, T.~W. Kephart, E.~W. Kolb, R.~Watkins and L.~M.
  Widrow, \emph{{Solutions to the strong CP problem in a world with gravity}},
  \href{http://dx.doi.org/10.1016/0370-2693(92)90491-L}{\emph{Phys. Lett.} {\bf
  B282} (1992) 132--136}, [\href{https://arxiv.org/abs/hep-ph/9203206}{{\tt
  hep-ph/9203206}}].

\bibitem{Kallosh:1995hi}
R.~Kallosh, A.~D. Linde, D.~A. Linde and L.~Susskind, \emph{{Gravity and global
  symmetries}}, \href{http://dx.doi.org/10.1103/PhysRevD.52.912}{\emph{Phys.
  Rev.} {\bf D52} (1995) 912--935},
  [\href{https://arxiv.org/abs/hep-th/9502069}{{\tt hep-th/9502069}}].

\bibitem{Banks:2010zn}
T.~Banks and N.~Seiberg, \emph{{Symmetries and Strings in Field Theory and
  Gravity}}, \href{http://dx.doi.org/10.1103/PhysRevD.83.084019}{\emph{Phys.
  Rev.} {\bf D83} (2011) 084019}, [\href{https://arxiv.org/abs/1011.5120}{{\tt
  1011.5120}}].

\bibitem{Susskind:1995da}
L.~Susskind, \emph{{Trouble for remnants}},
  \href{https://arxiv.org/abs/hep-th/9501106}{{\tt hep-th/9501106}}.

\bibitem{Saraswat:2016eaz}
P.~Saraswat, \emph{{Weak gravity conjecture and effective field theory}},
  \href{http://dx.doi.org/10.1103/PhysRevD.95.025013}{\emph{Phys. Rev.} {\bf
  D95} (2017) 025013}, [\href{https://arxiv.org/abs/1608.06951}{{\tt
  1608.06951}}].

\bibitem{Heidenreich:2016aqi}
B.~Heidenreich, M.~Reece and T.~Rudelius, \emph{{Evidence for a sublattice weak
  gravity conjecture}},
  \href{http://dx.doi.org/10.1007/JHEP08(2017)025}{\emph{JHEP} {\bf 08} (2017)
  025}, [\href{https://arxiv.org/abs/1606.08437}{{\tt 1606.08437}}].

\bibitem{Heidenreich:2017sim}
B.~Heidenreich, M.~Reece and T.~Rudelius, \emph{{The Weak Gravity Conjecture
  and Emergence from an Ultraviolet Cutoff}},
  \href{http://dx.doi.org/10.1140/epjc/s10052-018-5811-3}{\emph{Eur. Phys. J.}
  {\bf C78} (2018) 337}, [\href{https://arxiv.org/abs/1712.01868}{{\tt
  1712.01868}}].

\bibitem{Pal:2010bf}
S.~S. Pal, \emph{{Weak Gravity Conjecture, Central Charges and $\eta/s$}},
  \href{https://arxiv.org/abs/1003.0745}{{\tt 1003.0745}}.

\bibitem{Cheung:2014ega}
C.~Cheung and G.~N. Remmen, \emph{{Infrared Consistency and the Weak Gravity
  Conjecture}}, \href{http://dx.doi.org/10.1007/JHEP12(2014)087}{\emph{JHEP}
  {\bf 12} (2014) 087}, [\href{https://arxiv.org/abs/1407.7865}{{\tt
  1407.7865}}].

\bibitem{Harlow:2015lma}
D.~Harlow, \emph{{Wormholes, Emergent Gauge Fields, and the Weak Gravity
  Conjecture}}, \href{http://dx.doi.org/10.1007/JHEP01(2016)122}{\emph{JHEP}
  {\bf 01} (2016) 122}, [\href{https://arxiv.org/abs/1510.07911}{{\tt
  1510.07911}}].

\bibitem{Cottrell:2016bty}
G.~Shiu, P.~Soler and W.~Cottrell, \emph{{Weak Gravity Conjecture and Extremal
  Black Hole}},  \href{https://arxiv.org/abs/1611.06270}{{\tt 1611.06270}}.

\bibitem{Hebecker:2017uix}
A.~Hebecker and P.~Soler, \emph{{The Weak Gravity Conjecture and the Axionic
  Black Hole Paradox}},
  \href{http://dx.doi.org/10.1007/JHEP09(2017)036}{\emph{JHEP} {\bf 09} (2017)
  036}, [\href{https://arxiv.org/abs/1702.06130}{{\tt 1702.06130}}].

\bibitem{Lee:2018urn}
S.-J. Lee, W.~Lerche and T.~Weigand, \emph{{Tensionless Strings and the Weak
  Gravity Conjecture}},
  \href{http://dx.doi.org/10.1007/JHEP10(2018)164}{\emph{JHEP} {\bf 10} (2018)
  164}, [\href{https://arxiv.org/abs/1808.05958}{{\tt 1808.05958}}].

\bibitem{Andriolo:2018lvp}
S.~Andriolo, D.~Junghans, T.~Noumi and G.~Shiu, \emph{{A Tower Weak Gravity
  Conjecture from Infrared Consistency}},
  \href{http://dx.doi.org/10.1002/prop.201800020}{\emph{Fortsch. Phys.} {\bf
  66} (2018) 1800020}, [\href{https://arxiv.org/abs/1802.04287}{{\tt
  1802.04287}}].

\bibitem{Hamada:2018dde}
Y.~Hamada, T.~Noumi and G.~Shiu, \emph{{Weak Gravity Conjecture from Unitarity
  and Causality}},  \href{https://arxiv.org/abs/1810.03637}{{\tt 1810.03637}}.

\bibitem{Bonnefoy:2018mqb}
Q.~Bonnefoy, E.~Dudas and S.~Lust, \emph{{On the weak gravity conjecture in
  string theory with broken supersymmetry}},
  \href{https://arxiv.org/abs/1811.11199}{{\tt 1811.11199}}.

\bibitem{Crisford:2017gsb}
T.~Crisford, G.~T. Horowitz and J.~E. Santos, \emph{{Testing the Weak Gravity -
  Cosmic Censorship Connection}},
  \href{http://dx.doi.org/10.1103/PhysRevD.97.066005}{\emph{Phys. Rev.} {\bf
  D97} (2018) 066005}, [\href{https://arxiv.org/abs/1709.07880}{{\tt
  1709.07880}}].

\bibitem{Horowitz:2019eum}
G.~T. Horowitz and J.~E. Santos, \emph{{Further evidence for the weak gravity -
  cosmic censorship connection}},  \href{https://arxiv.org/abs/1901.11096}{{\tt
  1901.11096}}.

\bibitem{Huang:2006hc}
Q.-G. Huang, M.~Li and W.~Song, \emph{{Weak gravity conjecture in the
  asymptotical dS and AdS background}},
  \href{http://dx.doi.org/10.1088/1126-6708/2006/10/059}{\emph{JHEP} {\bf 10}
  (2006) 059}, [\href{https://arxiv.org/abs/hep-th/0603127}{{\tt
  hep-th/0603127}}].

\bibitem{Li:2006jja}
M.~Li, W.~Song, Y.~Song and T.~Wang, \emph{{A Weak gravity conjecture for
  scalar field theories}},
  \href{http://dx.doi.org/10.1088/1126-6708/2007/05/026}{\emph{JHEP} {\bf 05}
  (2007) 026}, [\href{https://arxiv.org/abs/hep-th/0606011}{{\tt
  hep-th/0606011}}].

\bibitem{Huang:2007st}
Q.-G. Huang, \emph{{Weak Gravity Conjecture for the Effective Field Theories
  with N Species}},
  \href{http://dx.doi.org/10.1103/PhysRevD.77.105029}{\emph{Phys. Rev.} {\bf
  D77} (2008) 105029}, [\href{https://arxiv.org/abs/0712.2859}{{\tt
  0712.2859}}].

\bibitem{Cheung:2014vva}
C.~Cheung and G.~N. Remmen, \emph{{Naturalness and the Weak Gravity
  Conjecture}},
  \href{http://dx.doi.org/10.1103/PhysRevLett.113.051601}{\emph{Phys. Rev.
  Lett.} {\bf 113} (2014) 051601}, [\href{https://arxiv.org/abs/1402.2287}{{\tt
  1402.2287}}].

\bibitem{Nakayama:2015hga}
Y.~Nakayama and Y.~Nomura, \emph{{Weak gravity conjecture in the AdS/CFT
  correspondence}},
  \href{http://dx.doi.org/10.1103/PhysRevD.92.126006}{\emph{Phys. Rev.} {\bf
  D92} (2015) 126006}, [\href{https://arxiv.org/abs/1509.01647}{{\tt
  1509.01647}}].

\bibitem{Heidenreich:2015nta}
B.~Heidenreich, M.~Reece and T.~Rudelius, \emph{{Sharpening the Weak Gravity
  Conjecture with Dimensional Reduction}},
  \href{http://dx.doi.org/10.1007/JHEP02(2016)140}{\emph{JHEP} {\bf 02} (2016)
  140}, [\href{https://arxiv.org/abs/1509.06374}{{\tt 1509.06374}}].

\bibitem{Montero:2016tif}
M.~Montero, G.~Shiu and P.~Soler, \emph{{The Weak Gravity Conjecture in three
  dimensions}}, \href{http://dx.doi.org/10.1007/JHEP10(2016)159}{\emph{JHEP}
  {\bf 10} (2016) 159}, [\href{https://arxiv.org/abs/1606.08438}{{\tt
  1606.08438}}].

\bibitem{Palti:2017elp}
E.~Palti, \emph{{The Weak Gravity Conjecture and Scalar Fields}},
  \href{http://dx.doi.org/10.1007/JHEP08(2017)034}{\emph{JHEP} {\bf 08} (2017)
  034}, [\href{https://arxiv.org/abs/1705.04328}{{\tt 1705.04328}}].

\bibitem{Hod:2017uqc}
S.~Hod, \emph{{A proof of the weak gravity conjecture}},
  \href{http://dx.doi.org/10.1142/S0218271817420044}{\emph{Int. J. Mod. Phys.}
  {\bf D26} (2017) 1742004}, [\href{https://arxiv.org/abs/1705.06287}{{\tt
  1705.06287}}].

\bibitem{Fisher:2017dbc}
Z.~Fisher and C.~J. Mogni, \emph{{A Semiclassical, Entropic Proof of a Weak
  Gravity Conjecture}},  \href{https://arxiv.org/abs/1706.08257}{{\tt
  1706.08257}}.

\bibitem{Cheung:2018cwt}
C.~Cheung, J.~Liu and G.~N. Remmen, \emph{{Proof of the Weak Gravity Conjecture
  from Black Hole Entropy}},
  \href{http://dx.doi.org/10.1007/JHEP10(2018)004}{\emph{JHEP} {\bf 10} (2018)
  004}, [\href{https://arxiv.org/abs/1801.08546}{{\tt 1801.08546}}].

\bibitem{Montero:2018fns}
M.~Montero, \emph{{A Holographic Derivation of the Weak Gravity Conjecture}},
  \href{https://arxiv.org/abs/1812.03978}{{\tt 1812.03978}}.

\bibitem{Polchinski:2003bq}
J.~Polchinski, \emph{{Monopoles, duality, and string theory}},
  \href{http://dx.doi.org/10.1142/S0217751X0401866X}{\emph{Int. J. Mod. Phys.}
  {\bf A19S1} (2004) 145--156},
  [\href{https://arxiv.org/abs/hep-th/0304042}{{\tt hep-th/0304042}}].

\bibitem{Harlow:2018jwu}
D.~Harlow and H.~Ooguri, \emph{{Constraints on symmetry from holography}},
  \href{https://arxiv.org/abs/1810.05337}{{\tt 1810.05337}}.

\bibitem{Aharonov:1959fk}
Y.~Aharonov and D.~Bohm, \emph{{Significance of electromagnetic potentials in
  the quantum theory}},
  \href{http://dx.doi.org/10.1103/PhysRev.115.485}{\emph{Phys. Rev.} {\bf 115}
  (1959) 485--491}.

\bibitem{Alford:1988sj}
M.~G. Alford and F.~Wilczek, \emph{{Aharonov-Bohm Interaction of Cosmic Strings
  with Matter}},
  \href{http://dx.doi.org/10.1103/PhysRevLett.62.1071}{\emph{Phys. Rev. Lett.}
  {\bf 62} (1989) 1071}.

\bibitem{Krauss:1988zc}
L.~M. Krauss and F.~Wilczek, \emph{{Discrete Gauge Symmetry in Continuum
  Theories}}, \href{http://dx.doi.org/10.1103/PhysRevLett.62.1221}{\emph{Phys.
  Rev. Lett.} {\bf 62} (1989) 1221}.

\bibitem{Alford:1989ch}
M.~G. Alford, J.~March-Russell and F.~Wilczek, \emph{{Discrete Quantum Hair on
  Black Holes and the Nonabelian {Aharonov-Bohm} Effect}},
  \href{http://dx.doi.org/10.1016/0550-3213(90)90512-C}{\emph{Nucl. Phys.} {\bf
  B337} (1990) 695--708}.

\bibitem{Alford:1990pt}
M.~G. Alford, S.~R. Coleman and J.~March-Russell, \emph{{Disentangling
  nonAbelian discrete quantum hair}},
  \href{http://dx.doi.org/10.1016/S0550-3213(05)80042-2}{\emph{Nucl. Phys.}
  {\bf B351} (1991) 735--748}.

\bibitem{Dvali:2007hz}
G.~Dvali, \emph{{Black Holes and Large N Species Solution to the Hierarchy
  Problem}}, \href{http://dx.doi.org/10.1002/prop.201000009}{\emph{Fortsch.
  Phys.} {\bf 58} (2010) 528--536},
  [\href{https://arxiv.org/abs/0706.2050}{{\tt 0706.2050}}].

\bibitem{Dvali:2007wp}
G.~Dvali and M.~Redi, \emph{{Black Hole Bound on the Number of Species and
  Quantum Gravity at LHC}},
  \href{http://dx.doi.org/10.1103/PhysRevD.77.045027}{\emph{Phys. Rev.} {\bf
  D77} (2008) 045027}, [\href{https://arxiv.org/abs/0710.4344}{{\tt
  0710.4344}}].

\bibitem{Dvali:2008tq}
G.~Dvali, M.~Redi, S.~Sibiryakov and A.~Vainshtein, \emph{{Gravity Cutoff in
  Theories with Large Discrete Symmetries}},
  \href{http://dx.doi.org/10.1103/PhysRevLett.101.151603}{\emph{Phys. Rev.
  Lett.} {\bf 101} (2008) 151603}, [\href{https://arxiv.org/abs/0804.0769}{{\tt
  0804.0769}}].

\bibitem{Preskill:1990ty}
J.~Preskill, \emph{{Quantum hair}},
  \href{http://dx.doi.org/10.1088/0031-8949/1991/T36/028}{\emph{Phys. Scripta}
  {\bf T36} (1991) 258--264}.

\bibitem{Coleman:1991ku}
S.~R. Coleman, J.~Preskill and F.~Wilczek, \emph{{Quantum hair on black
  holes}}, \href{http://dx.doi.org/10.1016/0550-3213(92)90008-Y}{\emph{Nucl.
  Phys.} {\bf B378} (1992) 175--246},
  [\href{https://arxiv.org/abs/hep-th/9201059}{{\tt hep-th/9201059}}].

\bibitem{Maldacena:2001ss}
J.~M. Maldacena, G.~W. Moore and N.~Seiberg, \emph{{D-brane charges in
  five-brane backgrounds}},
  \href{http://dx.doi.org/10.1088/1126-6708/2001/10/005}{\emph{JHEP} {\bf 10}
  (2001) 005}, [\href{https://arxiv.org/abs/hep-th/0108152}{{\tt
  hep-th/0108152}}].

\bibitem{Adler:1980pg}
S.~L. Adler, \emph{{A Formula for the Induced Gravitational Constant}},
  \href{http://dx.doi.org/10.1016/0370-2693(80)90478-5}{\emph{Phys. Lett.} {\bf
  B95} (1980) 241}.

\bibitem{Zee:1981mk}
A.~Zee, \emph{{Calculating Newton's Gravitational Constant in Infrared Stable
  {Yang-Mills} Theories}},
  \href{http://dx.doi.org/10.1103/PhysRevLett.48.295}{\emph{Phys. Rev. Lett.}
  {\bf 48} (1982) 295}.

\bibitem{Dowker:1991qe}
F.~Dowker, R.~Gregory and J.~H. Traschen, \emph{{Euclidean black hole
  vortices}}, \href{http://dx.doi.org/10.1103/PhysRevD.45.2762}{\emph{Phys.
  Rev.} {\bf D45} (1992) 2762--2771},
  [\href{https://arxiv.org/abs/hep-th/9112065}{{\tt hep-th/9112065}}].

\bibitem{GarciaGarcia:2018tua}
I.~Garcia~Garcia, \emph{{Properties of Discrete Black Hole Hair}},
  \href{https://arxiv.org/abs/1809.03527}{{\tt 1809.03527}}.

\bibitem{Preskill:1990bm}
J.~Preskill and L.~M. Krauss, \emph{{Local Discrete Symmetry and Quantum
  Mechanical Hair}},
  \href{http://dx.doi.org/10.1016/0550-3213(90)90262-C}{\emph{Nucl. Phys.} {\bf
  B341} (1990) 50--100}.

\bibitem{Aryal:1986sz}
M.~Aryal, L.~H. Ford and A.~Vilenkin, \emph{{Cosmic Strings and Black Holes}},
  \href{http://dx.doi.org/10.1103/PhysRevD.34.2263}{\emph{Phys. Rev.} {\bf D34}
  (1986) 2263}.

\bibitem{Achucarro:1995nu}
A.~Achucarro, R.~Gregory and K.~Kuijken, \emph{{Abelian Higgs hair for black
  holes}}, \href{http://dx.doi.org/10.1103/PhysRevD.52.5729}{\emph{Phys. Rev.}
  {\bf D52} (1995) 5729--5742},
  [\href{https://arxiv.org/abs/gr-qc/9505039}{{\tt gr-qc/9505039}}].

\bibitem{Reece:2018zvv}
M.~Reece, \emph{{Photon Masses in the Landscape and the Swampland}},
  \href{https://arxiv.org/abs/1808.09966}{{\tt 1808.09966}}.

\bibitem{Vafa:2005ui}
C.~Vafa, \emph{{The String landscape and the swampland}},
  \href{https://arxiv.org/abs/hep-th/0509212}{{\tt hep-th/0509212}}.

\bibitem{Ooguri:2006in}
H.~Ooguri and C.~Vafa, \emph{{On the Geometry of the String Landscape and the
  Swampland}},
  \href{http://dx.doi.org/10.1016/j.nuclphysb.2006.10.033}{\emph{Nucl. Phys.}
  {\bf B766} (2007) 21--33}, [\href{https://arxiv.org/abs/hep-th/0605264}{{\tt
  hep-th/0605264}}].

\bibitem{Adams:2006sv}
A.~Adams, N.~Arkani-Hamed, S.~Dubovsky, A.~Nicolis and R.~Rattazzi,
  \emph{{Causality, analyticity and an IR obstruction to UV completion}},
  \href{http://dx.doi.org/10.1088/1126-6708/2006/10/014}{\emph{JHEP} {\bf 10}
  (2006) 014}, [\href{https://arxiv.org/abs/hep-th/0602178}{{\tt
  hep-th/0602178}}].

\bibitem{Ooguri:2016pdq}
H.~Ooguri and C.~Vafa, \emph{{Non-supersymmetric AdS and the Swampland}},
  \href{http://dx.doi.org/10.4310/ATMP.2017.v21.n7.a8}{\emph{Adv. Theor. Math.
  Phys.} {\bf 21} (2017) 1787--1801},
  [\href{https://arxiv.org/abs/1610.01533}{{\tt 1610.01533}}].

\bibitem{Montero:2017yja}
M.~Montero, A.~M. Uranga and I.~Valenzuela, \emph{{A Chern-Simons Pandemic}},
  \href{http://dx.doi.org/10.1007/JHEP07(2017)123}{\emph{JHEP} {\bf 07} (2017)
  123}, [\href{https://arxiv.org/abs/1702.06147}{{\tt 1702.06147}}].

\bibitem{Danielsson:2018ztv}
U.~H. Danielsson and T.~Van~Riet, \emph{{What if string theory has no de Sitter
  vacua?}}, \href{http://dx.doi.org/10.1142/S0218271818300070}{\emph{Int. J.
  Mod. Phys.} {\bf D27} (2018) 1830007},
  [\href{https://arxiv.org/abs/1804.01120}{{\tt 1804.01120}}].

\bibitem{Obied:2018sgi}
G.~Obied, H.~Ooguri, L.~Spodyneiko and C.~Vafa, \emph{{De Sitter Space and the
  Swampland}},  \href{https://arxiv.org/abs/1806.08362}{{\tt 1806.08362}}.

\bibitem{delaFuente:2014aca}
A.~de~la Fuente, P.~Saraswat and R.~Sundrum, \emph{{Natural Inflation and
  Quantum Gravity}},
  \href{http://dx.doi.org/10.1103/PhysRevLett.114.151303}{\emph{Phys. Rev.
  Lett.} {\bf 114} (2015) 151303}, [\href{https://arxiv.org/abs/1412.3457}{{\tt
  1412.3457}}].

\bibitem{Brown:2015iha}
J.~Brown, W.~Cottrell, G.~Shiu and P.~Soler, \emph{{Fencing in the Swampland:
  Quantum Gravity Constraints on Large Field Inflation}},
  \href{http://dx.doi.org/10.1007/JHEP10(2015)023}{\emph{JHEP} {\bf 10} (2015)
  023}, [\href{https://arxiv.org/abs/1503.04783}{{\tt 1503.04783}}].

\bibitem{Ibanez:2017kvh}
L.~E. Ibanez, V.~Martin-Lozano and I.~Valenzuela, \emph{{Constraining Neutrino
  Masses, the Cosmological Constant and BSM Physics from the Weak Gravity
  Conjecture}}, \href{http://dx.doi.org/10.1007/JHEP11(2017)066}{\emph{JHEP}
  {\bf 11} (2017) 066}, [\href{https://arxiv.org/abs/1706.05392}{{\tt
  1706.05392}}].

\bibitem{Ibanez:2017oqr}
L.~E. Ibanez, V.~Martin-Lozano and I.~Valenzuela, \emph{{Constraining the EW
  Hierarchy from the Weak Gravity Conjecture}},
  \href{https://arxiv.org/abs/1707.05811}{{\tt 1707.05811}}.

\bibitem{Choi:2015fiu}
K.~Choi and S.~H. Im, \emph{{Realizing the relaxion from multiple axions and
  its UV completion with high scale supersymmetry}},
  \href{http://dx.doi.org/10.1007/JHEP01(2016)149}{\emph{JHEP} {\bf 01} (2016)
  149}, [\href{https://arxiv.org/abs/1511.00132}{{\tt 1511.00132}}].

\bibitem{Kaplan:2015fuy}
D.~E. Kaplan and R.~Rattazzi, \emph{{Large field excursions and approximate
  discrete symmetries from a clockwork axion}},
  \href{http://dx.doi.org/10.1103/PhysRevD.93.085007}{\emph{Phys. Rev.} {\bf
  D93} (2016) 085007}, [\href{https://arxiv.org/abs/1511.01827}{{\tt
  1511.01827}}].

\bibitem{Craig:2018yld}
N.~Craig and I.~Garcia~Garcia, \emph{{Rescuing Massive Photons from the
  Swampland}}, \href{http://dx.doi.org/10.1007/JHEP11(2018)067}{\emph{JHEP}
  {\bf 11} (2018) 067}, [\href{https://arxiv.org/abs/1810.05647}{{\tt
  1810.05647}}].

\bibitem{Lust:2017wrl}
D.~Lust and E.~Palti, \emph{{Scalar Fields, Hierarchical UV/IR Mixing and The
  Weak Gravity Conjecture}},
  \href{http://dx.doi.org/10.1007/JHEP02(2018)040}{\emph{JHEP} {\bf 02} (2018)
  040}, [\href{https://arxiv.org/abs/1709.01790}{{\tt 1709.01790}}].

\bibitem{Adelberger:2009zz}
E.~G. Adelberger, J.~H. Gundlach, B.~R. Heckel, S.~Hoedl and S.~Schlamminger,
  \emph{{Torsion balance experiments: A low-energy frontier of particle
  physics}}, \href{http://dx.doi.org/10.1016/j.ppnp.2008.08.002}{\emph{Prog.
  Part. Nucl. Phys.} {\bf 62} (2009) 102--134}.

\bibitem{Wagner:2012ui}
T.~A. Wagner, S.~Schlamminger, J.~H. Gundlach and E.~G. Adelberger,
  \emph{{Torsion-balance tests of the weak equivalence principle}},
  \href{http://dx.doi.org/10.1088/0264-9381/29/18/184002}{\emph{Class. Quant.
  Grav.} {\bf 29} (2012) 184002}, [\href{https://arxiv.org/abs/1207.2442}{{\tt
  1207.2442}}].

\bibitem{Ibanez:2017vfl}
L.~E. Ibanez and M.~Montero, \emph{{A Note on the WGC, Effective Field Theory
  and Clockwork within String Theory}},
  \href{http://dx.doi.org/10.1007/JHEP02(2018)057}{\emph{JHEP} {\bf 02} (2018)
  057}, [\href{https://arxiv.org/abs/1709.02392}{{\tt 1709.02392}}].

\end{thebibliography}\endgroup

\end{document}